%Paper: 9109028
%From: KALYAN%tifrvax.bitnet@pucc.PRINCETON.EDU
%Date: Wed, 18 Sep 91 15:28 IST

------------------------------------------------------------------------------

%%%%%%%%%%%%%%%%%%
%%%%%%%%%%%%
%%%   PROCESS THIS PAPER WITH PHYZZX
%%%%%%%%%%%%
%%%%%%%%%%%%%%%%%%
\input phyzzx

\Pubnum {TIFR/TH/91--41}
\date {September 1991}
\titlepage
%\title { New Special Operators in $W -$gravity Theories }
\voffset=24pt
\title { NEW SPECIAL OPERATORS IN $W -$GRAVITY THEORIES }
\author{S. Kalyana Rama \foot{e-mail : kalyan@tifrvax.bitnet}}
\address {Tata Institute of Fundamental Research,\break
Homi Bhabha Road, Bombay 400 005, INDIA}
\abstract {
We find new special physical operators of $W_3 -$gravity having non
trivial ghost sectors.  Some of these operators may be viewed as the
liouville dressings of the energy operator of the Ising model coupled to
{\it 2d~gravity} and this fills in a gap in the connection between pure
$W_3 -$gravity and Ising model coupled to 2d gravity found in our previous
work.  We formulate a selection rule required for the calculation of
correlators in $W -$gravity theories. Using this rule, we construct the
non ghost part of the new operators of $W_N -$gravity and find that they
represent  the $(N , N+1)$ minimal model operators from both inside and
outside the minimal table. Along the way we obtain the canonical spectrum
of $W_N -$gravity for all $N$ .  }
\vfill

%\centerline {(Submitted to )}

\endpage

\NPrefs
\def\define#1#2\par{\def#1{\Ref#1{#2}\edef#1{\noexpand\refmark{#1}}}}
\def\con#1#2\noc{\let\?=\Ref\let\<=\refmark\let\Ref=\REFS
         \let\refmark=\undefined#1\let\Ref=\REFSCON#2
         \let\Ref=\?\let\refmark=\<\refsend}

\define\Z
A. B. Zamolodchikov, Teo. Mat. Fiz 65 (1985) 347.

\define\FZ
V. A. Fateev and A. B. Zamolodchikov, Nucl. Phys. B280 (1987) 644.

\define\FL
V. A. Fateev and S. L. Lykyanov, Int. J. Mod. Phys. A3 (1988) 507.

\define\Wgang
For example,
F. Bais et al, Nucl. Phys. B304 (1988) 348;
A. Bilal and J. L. Gervais, Nucl. Phys. B326 (1989) 222
and references therein;
I. Bakas, Comm. Math. Phys. 123 (1989) 627;
M. Bershadsky and H. Ooguri, Comm. Math. Phys. 126 (1989) 49;
C. M. Hull, Phys. Lett. B240 (1990) 110;
K. Schoutens et al,  Nucl. Phys. B349 (1991) 791;
E. Bergshoeff et al, Phys. Lett. B243 (1990) 350;
Y. Matsuo, Phys. Lett. B227 (1989) 209;
L. Romans,  Nucl. Phys. B352 (1991) 829.

\define\DDR
S. R. Das, A. Dhar and S. K. Rama,
Preprint TIFR/TH/91--11, (February 1991), to appear in Mod. Phys. Lett. A ;
Preprint TIFR/TH/91--20, (March 1991), to appear in Int. J. Mod. Phys. A.

\define\FKN
M. Fukuma, H. Kawai and R. Nakayama, Int. J. Mod. Phys. A6 (1991) 1385;
R. Dijkgraaf, E. Verlinde and H. Verlinde,  Nucl. Phys. B348 (1991) 435;
J. Goeree,  Nucl. Phys. B358 (1991) 737.

\define\Doug
M. Douglas, Phys. Lett. B238 (1990) 176;
E. Witten, Nucl. Phys. B340 (1990) 281;
J. Distler, Nucl. Phys. B342 (1990) 523;
E. verlinde and H. Verlinde, Preprint PUPT--1176 (1990);
R. Dijkgraaf and E. Witten, Nucl. Phys. B342 (1990) 486;
E. Witten, preprint IASSNS-HEP-90/37 (1990).

\define\D
V. S. Dotsenko, Preprint PAR--LPTHE 91--18, (February 1991).

\define\K
Y. Kitazawa, Phys. Lett. B265 (1991) 262.

\define\discrete
For example,
D. J. Gross, I. Klebanov  and M. Newman, Nucl. Phys. B350 (1990) 621;
A. M. Polyakov, Mod. Phys. Lett. A6 (1991) 635;
U. Danielsson and D. J. Gross, Preprint PUPT--1258 (1991);
S. R. Das, A. Dhar, G. Mandal and S. R. Wadia, TIFR preprint to appear;
P. Bouwknegt, J. McCarthy and K. Pilch, Preprint CERN--TH-6162/91;
G. Moore and N. Seiberg, Preprint RU--91--29 / YCTP--P19--91.

\define\W
E. Witten,  Preprint IASSNS-HEP-91/51 (1991);
J. Distler and P. Nelson, Preprint UPR--462T / PUPT--1262 (August 1991);
S. Mukhi, S. Mukherji and A. Sen,  Preprint TIFR/Th/91--25 (May 1991);
C. Imbimbo, S. Mahapatra and S. Mukhi,
Preprint TIFR/Th/91--27 / GEF--TH 8/91 (May 1991).

\define\LZ
B. H. Lian and G. J. Zuckerman, Phys. Lett. B254 (1991) 417 ; ibid B266
(1991) 21.

\define\FQS
For example,
D. Friedan, Z. Qiu and S. Shenker, Phys. Rev. Lett. 52 (1984) 1575.

\define\Thier
J. Thierry-Mieg, Phys. Lett. B197 (1987) 368.

\define\Bouwk
P. Bouwknegt, private communication;
J. B. Zuber, private communication.

\define\Dh
A. Dhar, private communication.

\define\FMS
For example,
D. Friedan, E. Martinec  and S. Shenker, Nucl. Phys. B271 (1986) 93.

\define\GL
M. Goulian and M. Li, Phys. Rev. Lett. 66 (1991) 2051;
P. Di Francesco and D. Kutasov, Phys. Lett. B261 (1991) 385;
N. Sakai and Y. Tanii, Preprint TIT/HEP--168, STUPP--91--116 (March 1991).

\centerline{1. INTRODUCTION}

$W_N -$algebras \Z\FL\ , are non-linear
extensions of Virasoro algebra by higher spin currents with spins $s, \,
2 \leq s \leq N , \,$ the spin -- 2 current being the stress tensor, $T$ .
$W -$gravities are obtained by gauging $W -$algebras and their
formal aspects have been studied in \FL\FZ\Wgang\
while the physical properties and the spectrum of $W -$gravities and
$W -$strings have been studied in detail in \DDR\ for $W_3 $ and $W_4$ .

In \DDR\ , among other things we obtained the ``canonical" physical
operators of pure $W_N -$gravity, for $N = 3, 4$ , and found an intriguing
connection with unitary minimal models coupled to 2d gravity. We found
that all the canonical physical operators of $W_N -$gravity
can be regarded as the ``liouville" dressings of the diagonal operators of
the $(N , N+1)$ minimal models coupled to 2d gravity, for $N = 3, 4$ ,
and conjectured that this phenomenon is true for
all $N $. However we were unable to represent
the non diagonal operators in the context of $W -$gravity.
Also, recent studies \D\K\ indicate that the minimal model
operators from outside the minimal table do not decouple when
coupled to gravity. But the representation of such operators in the
$W -$gravity framework is not known either.

Connections such as above have appeared elsewhere too. The double scaling
limit of the multimatrix models is believed
to be $(p,q)$ minimal models coupled to gravity  and the $W_k -$constraints
appear \FKN\ in the multimatrix models as Dyson -- Schwinger
equations. Also, in the topological
field theoretic description of Ising model coupled to topological gravity,
the diagonal operators are special and appear as topological matter
primaries while the energy operator appears as ``gravitational
descendent" .

In view of such discoveries, it is important to know how to represent the
non diagonal operators of the minimal model and also the operators from
outside the minimal table in the framework of $W -$gravity theories.
Knowing this is crucial for further understanding of $W -$gravity and
$W_k -$constraints of multimatrix models.

In this paper, we describe the representation of such operators for Ising
model in the context of $W_3 -$gravity. To be precise, we show how to
represent the energy operator of the Ising model as new physical operators
in the $W_3 -$gravity spectrum. These operators have non trivial ghost
sectors and, as an added bonus, they also represent another Ising model
operator, from outside the minimal table.

We further formulate a selection rule, required for the calculation of
correlators in $W -$gravity theories. Using this rule, we construct the
non ghost part of the new special operators in $W_N -$gravity. They are
found to represent the $(N , N+1)$ minimal model operators from both inside
and outside the minimal table.  (The construction of
the non trivial ghost sectors for these new operators requires
the BRST charge Q$_B$ for $W_N -$gravity, which is not known for $N > 3$).

Along the way we obtain the canonical physical operators of
$W_N -$gravity, with the standard ghost sector, for all $N$
and find that they reprsent the diagonal operators of
the $(N , N+1)$ minimal model coupled to 2d gravity. We also show that these
canonical physical operators can be expressed as the non singular
composites of the screening charges, a fact first discovered in \DDR\ for
$N = 3, 4$ and conjectured for all $N$ .

\vskip 1.0cm

\centerline{ 2. $W -$GRAVITY SPECTRUM }

In this section we briefly review the relevent aspects of \DDR\FL\ and
give the construction of the $W_N -$gravity spectrum for all $N$ .
A free field representation of $W -$algebra
can be obtained in terms of (N -- 1) free scalar fields
$ \vec \phi = (\phi_1, \phi_2, \ldots , \phi_{N-1}) $ obeying the OPEs
$\phi_a(z) \phi_b(0) = - 2 \delta_{ab} log(z) . $ (In what follows, all the
vectors are (N -- 1) dimensional vectors). We describe the construction of
$W -$currents using the weights ${\vec h}_k , $ the simple roots
$ {\vec e}_k = {\vec h}_k - {\vec h}_{k+1} $ and the fundamental
weights ${\vec \omega}_k = \sum_{m=1}^k {\vec h}_m $ of
the representation of SU(N); here $ k = 1,2, \ldots , N-1 . $ We further
define a vector ${\vec h}_N $ by $ \sum_{k=1}^N {\vec h}_k = 0 . $
Furthermore the Weyl vector $\vec \rho$ is given by
$\vec \rho = \sum_{m=1}^{N-1} {\vec \omega}_m . $
While the choice of ${\vec h}_m$ does not matter, we often use the
canonical choice :
$$\eqalign{
& {\vec h}_1 = ( {1 \over \sqrt{2}} , {1 \over \sqrt{6}}, \cdots,
{1 \over \sqrt{N(N-1)}} ) \;\;,\;\;
{\vec h}_2 = ( {-1 \over \sqrt{2}} , {1 \over \sqrt{6}}, \cdots,
{1 \over \sqrt{N(N-1)}} )  \cr
& {\vec h}_3 = ( 0, {-2  \over \sqrt{6}}, \cdots,
{1 \over \sqrt{N(N-1)}} )  \;\; \ldots  \;\;
{\vec h}_N = ( 0,0, \cdots, {1 \over \sqrt{N(N-1)}} )     } \eqn\two
$$
With this choice the Weyl vector $\vec \rho$ becomes
$ \vec \rho =  { 1 \over 2 } ( \sqrt{2}, \sqrt{6}, \cdots, \sqrt{N(N+1)} ) . $
Note also that $\rho^2 = { {N(N^2-1)} \over {12} } . $

Now we define the quantities $U_k $ through the quantum Miura
transformation
$$
\prod_{m=1}^N  (\partial + { {{\vec h}_m \cdot \partial \vec \phi} \over
{2 i \alpha_0} }  ) = \sum_{k=0}^N U_k (z) \partial^{N-k}      \eqn\three
$$
where $\partial = { \partial \over {\partial z} } $ and $\alpha_0 $ is a free
parameter. Then, as explained in \DDR\ , the spin -- k currents $W_k(z) $
are given in terms of $U_m(z), m \leq k , $ and their derivatives upto an
overall constant which is fixed by requiring the central terms to be in
the canonical form. Thus for example, the stress tensor $ T ( \equiv W_2 ) $
is given by
$$
 T = (i \alpha_0 \sqrt{2})^2 U_2(z) = - { 1 \over 4 }
( \partial \vec \phi )^2
 + i \alpha_0 \vec \rho \cdot \partial^2 \vec \phi  . \eqn\four
$$
The modes of the $W_k$ currents $(k = 2, \cdots, N ) , $ defined by
$W_k(z) = \sum { {W_k(n)} \over {z^{n+k}} } , L_n \equiv W_2 (n)$,
form the $W_N -$algebra with the central charge
$$
C(N) = (N - 1) (1 - 2 \alpha_0^2 N(N+1))  . \eqn\five
$$
In $W -$gravity, for each $W_k$ , one has the ghost  fields $(b_k, c_k) $
of dimensions $(k, 1-k) $ respectively. Thus the central charge of the
entire ghost system is given by
$$
C_{gh}(N) = \sum_{k=2}^N -2 (1 + 6 k^2 - 6 k) =
- (N - 1) (4 N^2 + 4 N + 2 ) .   \eqn\six
$$
The critical central charge is obtained by requiring $C(N) + C_{gh}(N) = 0 $,
which gives
$$
\alpha_0^2 = - { {(2 N + 1)^2} \over {2 N (N + 1)} } .  \eqn\seven
$$
For  future use, we further define the quantities $\alpha_{\pm}$ by
$$
\alpha_{\pm} = { 1 \over 2 } ( \alpha_0 \pm \sqrt{\alpha_0^2 + 2} ) .
\eqn\eight $$
In $W -$gravity the ``canonical''
physical states (BRST closed but not BRST exact) are of the form
$$
\vert \Psi >_{phys} = \vert matter > \otimes \vert 0 >_{gh} \eqn\nine
$$
upto BRST exact terms, where the ghost vacuum is given by
$$
\vert 0 >_{gh} = \prod_{k=2}^N \prod _{n=1}^{k-1} c_k(n) \vert 0 > .
\eqn\ten $$
The ``canonical" physical states $\vert \Psi >_{phys} $ are assigned a
ghost number zero.
{}From above, one gets the $L_0 -$intercept for the matter sector as
$$
 L_0 \vert matter > = { {N(N^2 -1)} \over 6 } \vert matter > = ( 2 \rho^2 )
 \vert matter > . \eqn\eleven
$$
The other $W_k (0) , (k > 2), $ intercepts depend on the detailed structure
of the BRST charge.

The non ghost part of the operators which create the physical state are
of the form
$$ V_{\beta} (z) = e^{ i \vec \beta \cdot \vec \phi } . \eqn\twelve $$
The vector $\vec \beta $ can be determined from the
$W_k(0) , (2 \leq k \leq N) $ intercepts. However, since
this involves a knowledge of the BRST charge Q$_B$ which is not known for
$N > 3 $, we shall adopt
a correspondence principle given in \DDR\ to derive the $W -$gravity
spectrum. This principle briefly means the following. When $ \vec \beta =
{\vec \beta}_0 \equiv a \alpha_0 \vec \rho $, $a$ any constant,
the exponent ${\vec \beta}_0 \cdot \vec \phi $ in $ V_{\beta_0} $
is precisely in the combination which appears as the background
charge term in the stress tensor T in \four\ . For this and other reasons
\DDR\ , these are referred to as ``cosmological constant" operators.
Moreover, this operator is a solution for $W_3 -$gravity for any value of
$\alpha_0$  and for $W_4 -$gravity in the classical limit when $ \alpha_0^2
 \rightarrow - \infty $ . We assume that a correspondence principle holds
whereby the cosmological constant operator will continue to be a solution
for $W_N -$gravity for any value of $N $ and $\alpha_0$ .
The value of $ a $ in $ {\vec \beta}_0 = a \alpha_0 \vec \rho $
is determined by the $L_0 -$intercept
$( = 2 \rho^2 )$ in \eleven\ . One may use this solution to determine all the
$W_k (0) , (k > 2), $ intercepts and then solve for $\vec \beta$'s in
\twelve\ which have the same intercepts. They would constitute the
canonical physical spectrum of $W -$gravity.

In solving for $\vec \beta$  we note, following \FL\ ,
that by operating \three\
on $V_{\beta} (z) $ in \twelve\ and then operating the resulting equation on
functions of $z$ , say $z^j , j = 1,2, \ldots, N-2 , $ one can obtain the
following equations
$$
\prod_{m=1}^N ( N - m + j - {{{\vec h}_m \cdot  \vec \beta}
\over {\alpha_0}} )
= \sum_{k=0}^j { j! \over {(j - k)!} }  U_{N-k} (0) \eqn\thirteen
$$
which determine the intercepts $U_k (0) $, and hence $W_k (0) $.
Equation \thirteen\ has a permutation symmetry. Let ${\vec \beta}_0 $
corresponding to the set  $ \{ m_0  \} = (1,2, \ldots , N) $ be a solution
to \thirteen\ and let $ \vec \beta $ correspond to the set
$ \{ m \} = (m_1, m_2, \ldots , m_N) $ obtained by a
permutation of  $\{ m_0 \} $ . Then,
the left hand side of \thirteen\ is invariant under the
transformation ${\vec \beta}_0 \rightarrow {\vec \beta} $ and
$$
{\vec h}_{m_0} \cdot {\vec \beta}_0 + \alpha_0 m_0 =
{\vec h}_m \cdot \vec \beta + \alpha_0 m  . \eqn\fourteen
$$
Hence, $ \vec \beta $ satisfying \fourteen\ will be a new solution having
the same $W_k(0) $
intercepts as $\beta_0$ . Since, we know by the
correspondence principle that the cosmological constant operator,
$ {\vec \beta}_0 = a \alpha_0 \vec \rho $,  is a
solution of $W -$gravity constraints, we can obtain all the $ N! $
solutions by solving \fourteen\ for
${\vec \beta}$ corresponding to $\{ m \}$ .

To simplify further calculations, we define new parameters
$ i_k, \, b_k, \, K_0 $ and $ K_{\pm} $ by
$$ \eqalign{
& m_{i_k}  = k , \; m_{i_k} \in \{ m \} , \; k = 1, 2, \ldots , N  \cr
& \alpha_0  = { {i K_0} \over {\sqrt{2 N ( N + 1)}} }, \;\;
\alpha_{\pm} = { {i K_{\pm}} \over {\sqrt{2 N ( N + 1)}} } ,     \cr
& \vec \beta  = {i \over {\sqrt{2 N (N + 1)}} }
(b_1 \sqrt{2} , b_2 \sqrt{6} , \ldots , b_{N-1} \sqrt{N(N-1)} )
} \eqn\fifteen
$$
Note that from \seven\ and \eight\ ,
$K_0 = 2 N + 1 , K_+ = N + 1 ,K_- = N . $
However, we will not substitute these values in our expressions.
Requiring the dimension of $V_{\beta_0} $,
$ {\vec \beta}_0 = a \alpha_0 \vec \rho $, to be $2 \rho^2 $, we get
$$ 1 - a = \pm {1 \over K_0} \eqn\seventeen $$
where we have used the fact that the dimension of $V_{\beta} $ is
$ ({\vec  \beta}^2 - 2 \alpha_0 \vec \rho \cdot \vec \beta ) . $  We choose
the + sign in \seventeen\ . ( The other sign will correspond to
one of the solutions obtained from \fourteen\ ). Now solving  \fourteen\
for $\vec \beta$ corresponding to a given $\{ m \} $, parametrised as in
\fifteen\ , we obtain
$$ \eqalign{
b_1 + b_2 + \cdots + b_{N-1} + K_0 & = { {a K_0 (N + 1)} \over 2 }
+ (1 - a) i_1 K_0 \cr
- b_1 + b_2 + \cdots + b_{N-1} + 2 K_0 & = { {a K_0 (N + 1)} \over 2 }
+ (1 - a) i_2 K_0 \cr
- 2 b_2 + \cdots + b_{N-1} + 3 K_0 & = { {a K_0 (N + 1)} \over 2 }
+ (1 - a) i_3 K_0 \cr  & \cdots \cr
- (N - 1) b_{N-1} + N K_0 & = { {a K_0 (N + 1)} \over 2 }
+ (1 - a) i_N K_0  }\eqn\eighteen
$$
The solution of the above equations is given by
$$
b_p = ( { K_0 \over 2 } + { l_p \over {p(p + 1)} } ) \; , \;
p = 1,2, \ldots, N-1   \eqn\nineteen
$$
where $l_p $ is defined by
$$ l_p = i_1 + i_2 + \cdots + i_p - p i_{p+1} . \eqn\twenty $$
Equation \nineteen\ describes the canonical
physical spectrum of $W_N -$gravity for any $N$ .

Denoting the dimension of $e^{i \beta_p \phi_p}$ by
$\Delta_p ( = \beta_p^2 - 2 \alpha_0 \rho_p \beta_p ) $ , we get
$$
\Delta_p = { 1 \over  { (2 \rho_p)^2 2 N (N + 1) } } \;
           ( ({ { K_0 p (p + 1) } \over 2 })^2 - l_p^2 )    \eqn\twentyone
$$
where $ \rho_p = { { \sqrt{p (p + 1)} } \over 2 } $ is the $p$th component
of the Weyl vector $\vec \rho$ . In particular, the dimension of
$e^{i \beta_1 \phi_1} $, where $\phi_1 $ is the ``liouville mode" , is given
by
$$
\Delta_1 \equiv 1 - h_1 (N) = 1 - { {l_1^2 - 1} \over {4 N ( N + 1)} } .
\eqn\twentytwo
$$
{}From the definition of $l_1 $, it follows that
$ 1 \leq \vert l_1 \vert \leq N-1 $ . Comparing $h_1 (N) $ in \twentytwo\
with the dimensions $h_{p,q} $ of the $(p,q) $ minimal model operators
\FQS\ one sees that $e^{i \beta_1 \phi_1 } $ can be
interpreted as the ``liouville " dressing of the diagonal operators
of the $(N,N+1) $ minimal models, represented in $W_N -$gravity by the
fields $( \phi_k ; b_{k+1}, c_{k+1} ) , 2 \leq k < N $ .
Thus it follows that all the operators in the canonical
spectrum of $W_N -$gravity, for
any N, can be regarded as  the dressings of the diagonal operators of the
$(N,N+1) $ minimal models coupled to 2d gravity.

\vskip 1.0cm

\centerline{ 3. CONSTRUCTION OF ENERGY OPERATOR }

In this section we describe the representation of the energy operator of
the Ising model in the framework of $W_3 -$gravity.
Recently, Lian and Zuckerman have given in \LZ\ a cohomological
classification of physical states in $C \leq 1$ conformal matter coupled to
2d gravity and predicted new ``discrete" states with non trivial ghost
sectors. The discrete states have been the subject of several recent
studies \discrete\W\ and the physical consequences of such states with non
trivial ghost sectors have been examined by E. Witten \W\ in detail.
Some of these states with non trivial ghost sectors
have been constructed explicitly in \W\ .
The method of construction is essentially as follows: consider the
$(p,q)$ minimal models coupled to 2d gravity and the liouville operator
$e^{i \beta_L \phi_L} $ which dresses up a null state built over a
primary $\Phi_M$ of the minimal model. Then, reference \LZ\
guarantees that there exist physical states with non trivial ghost number
which can be built by acting upon
$e^{i \beta_L \phi_L} \Phi_M c_1 \vert 0 > $ with matter, liouville and
ghost oscillators. We adopt such a procedure for the case of
$W_3 -$gravity and construct explicitly new physical states with non
trivial ghost number.

We note that
for $W-$minimal models, no classification as in \LZ\ exists. However, we give
the following arguments to motivate our construction; the reasoning is
admittedly not rigorous, particulary due to the lack of understanding of
$W -$matter couplings to $W -$gravity. In the $W -$minimal models, there are
null states built over the primaries. For
example, in $W_3 -$minimal models \FZ\ , each primary state is labelled by
4 integers $(n,n';m,m')$ and it is known that null states over these
primaries exist at levels $nm$ and $n'm'$ \Bouwk\ . Now, if a $W -$minimal
model is coupled to $W -$gravity, one can obtain the $W -$gravity dressings
of such null states. Then analogous to the case of matter coupled to 2d
gravity, one can try to construct states with non trivial ghost number by
acting with oscillators on $W -$gravity dressings and the $W -$minimal
model primaries. Specialising to the case of the identity operator of the
$W -$minimal model with zero central charge, one should be able to obtain
the physical states in $W -$gravity with non trivial ghost  number.

Now we proceed as follows.
Using the formulas of \FZ\ , one discovers that a null state can be
built at level 2 over an identity operator of the $W -$minimal model with
zero central charge. This would mean that the $W_3 -$gravity dressing
should have dimension 2 (instead of the usual 4) and a ghost number $+1$ .
(The canonical physical states of section 2 are assigned a ghost number
zero). The knowledge of other
$W -$intercepts can be obtained by the correspondence principle.
Acting on such a state with required number of
oscillators, we find a physical state which is BRST closed but not BRST
exact.

We choose the ghost sector of the physical state  to be
$$
\vert 0 >_{gh} = c_2 (1) c_3 (0) c_3 (1) c_3 (2) \vert 0 >
\eqn\twentythree
$$
which is equivalent to \ten\ and, hence, also assigned
the same ghost number zero. We start with the operators of the form  %\break
$\Psi_{\beta} = e^{i \vec \beta \cdot \vec \phi}
c_2 (1) c_3 (0) c_3 (1) c_3 (2) $
where the dimension of $e^{i \vec \beta \cdot \vec \phi} $ is 2.
The BRST charge Q$_B$ is given in terms of modes in \Thier\ and in terms
of fields it is given by $Q_B = \oint d z j_B (z) $, where \Dh\
$$
j_B (z) = c_2 ( T + T(b_3,c_3) + {1 \over 2} T(b_2,c_2) ) + c_3 W_3
+ { 8 \over {261} } T b_2 c_3 \partial c_3
- { {25} \over {522} } b_2 \partial c_3 \partial^2 c_3
+ { {25} \over {783} } b_2 c_3 \partial^3 c_3  .   \eqn\twentyfour
$$
In the above expression $T(b_k, c_k) $ is the standard stress tensor
\FMS\ for the spin -- k ghost fields.  The operators $\Psi_{\beta}$
given above are not BRST closed. However, we find that the operators given by
$$ \tilde{\Psi}_{\beta} = c_2 (-2) \Psi_{\beta} ,  \eqn\twentyfive $$
where $\vec \beta$s are such that
the dimension of $e^{i \vec \beta \cdot \vec \phi} $ is 2,
are BRST closed but not BRST exact and hence, are genuine physical states
with ghost number $+1$ .

Now using the methods of section 2, one can find all the solutions of
$\vec \beta$ that appear in \twentyfive\ . The dimension of
the ``liouville field" $e^{i \beta_1 \phi_1} $ turns out to be
$\Delta_1 = 1 - h_1 (3) $ where
$$ h_1 (3) = { 1 \over 2 } \;\; , \;\; { {33} \over {16} } .
\eqn\twentysix  $$
Thus we see that all the physical operators $\tilde{\Psi}_{\beta} $,
which have a non
trivial ghost number $+ 1$, can be interpreted as
the liouville dressings of either the energy operator, $\Phi_{2,1}$ or the
operator $\Phi_{4,2} $ of the Ising model coupled to 2d gravity. Note that
the second operator $\Phi_{4,2} $ is from outside
the minimal table. However, it is not
surprising that this operator appears in the theory, in fact it is rather
desirable, in the light of the discovery of \D\K\ that operators from outside
the minimal table fail to decouple when the minimal models are coupled to
gravity. The decoupling in \D\K\ fails
because of the poles arising from the liouville sector which
precisely cancel the zeroes of the matter sector correlators involving
operators from outside the minimal table.

Thus we have explicitly constructed in $W_3 -$gravity  new special
physical operators with
non trivial ghost sectors which can be interpreted as the liouville
dressings of the operators of the Ising model from both inside and outside
the minimal table. Moreover, the physical operators $\tilde{\Psi}_{\beta} $
also satisfy another selection rule described in section 4.

A similar analysis of $W_N -$gravity for  $N > 3$, is not possible at present
since the corresponding BRST charge, Q$_B$ is not known.
However, it is quite likely that for any $N$, there exist special
operators similar to the ones constructed here satisfying the selection rule
mentioned above. These operators, as will be shown in the next section,
correspond to the minimal model operators from both inside
and outside the minimal table coupled to 2d gravity.

\vskip 1.0cm

\centerline{ 4. SCREENING CHARGES AND SELECTION RULE }

There has been some progress recently \D\K\GL\ in calculating the
three point functions of $ c \leq 1 $ matter coupled to 2d gravity. The
main ingredient in these works is the idea of analytically continuing $s $,
the number of screening charges that appear in the correlation functions
after integrating out the zero modes
of the liouville field, from positive integer values to
fractional \K\GL\ or negative integer \D\ values. In the
approach of \K\GL\ a semiclassical correspondence
principle is used to select the screening charge term in the
liouville action and hence after integrating out
the liouville zero mode, only one type of screening charge appears . This
necessitates the analytic continuation of $s$ into fractional values. In the
more natural approach of \D\
no semiclassical correspondence principle is used and both types of
screening charges appear  after doing the liouville sector functional integral
perturbatively. Because of this it is sufficient to anlytically
continue $s$ into negative integers only.

The screening charges in $W_N -$gravity are
$ S_k^{\pm} = e^{i \alpha_{\pm} {\vec e}_k \cdot \vec \phi} \; , \;
k = 1,2, \ldots, N-1  $.
However, no semiclassical correspondence principle is
known for $W_N -$gravity
and hence it is natural to include all the screening charge terms
in the action. Thus in calculating the
correlation functions $< \prod_{i=1}^L V_{\beta_i} > $
in $W -$gravity, one would obtain
the following condition a la \D\ :
$$
\sum_{i=1}^L {\vec \beta}_i + \sum_{m=1}^{N-1} (\alpha_+ {\cal E}_m^+
+ \alpha_- {\cal E}_m^- ) {\vec e}_m = 2 \alpha_0 \vec \rho .
\eqn\twentyeight
$$
Since we have included all the screening charge terms we expect to
analytically continue ${\cal E}_m^{\pm} $ into negative integers only and
not into fractional values. In what follows we impose it as a condition and
require ${\cal E}_m^{\pm}$ to take integer values only.
Since this requirement has to hold good for any arbitrary
L and for any $\vec \beta$ in the $W -$gravity spectrum and since
$ 2 \alpha_0 \vec \rho = \sum_{m=1}^{N-1} m (N - m) (\alpha_+ + \alpha_- )
{\vec e}_m $ , \twentyeight\ implies that
the $\vec \beta$'s must be expressible as
$$
\vec \beta =  \sum_{m=1}^{N-1} (\alpha_+ \epsilon_m^+
+ \alpha_- \epsilon_m^- ) {\vec e}_m         \eqn\twentynine
$$
where $\epsilon^{\pm}$ are positive or negative integers only.
This imposes a selection rule on $\vec \beta$ and hence on the physical
operators of $W -$gravity. Note that the canonical physical operators in
section 2 and the new physical operators
$\tilde{\Psi}_{\beta} $ in section 3 all satisfy this rule.

We now proceed to construct other new operators in the $W -$gravity
theory. Since we have neither the BRST charge Q$_B$ , nor the analog of
Lian-Zuckerman theorem we proceed as follows. In general,
the physical states $\tilde{\Psi}_{\beta} $
with non trivial ghost numbers have total dimension zero and
are produced by acting with  oscillators on
operators of the type $\Psi_{\beta} = e^{i \vec \beta \cdot \vec \phi}
\prod_{k=2}^N \prod_{n=1}^{k-1} c_k (n) $ (note that the states obtained
by acting with
$c_k(0) , 2 < k \leq N $ on $\Psi_{\beta}$
are equivalent to $\Psi_{\beta} $
and, hence, can be taken as a starting point instead of $\Psi_{\beta} $
and assigned the same ghost number zero ) .
Hence, we consider
$\vec \beta$ such that the dimension $\Delta $
of $e^{i \vec \beta \cdot \vec \phi} $ is given by
$$ \Delta = 2 \rho^2 - G  \eqn\thirty $$
where $G$ is a non negative integer. We then impose \twentynine\ and
obtain the special operators of $W -$gravity by using the methods of
section 2.

We have the cosmological constant operator
${\vec \beta}_0 = a \alpha_0 \vec \rho $ as a solution by the
correspondence principle. From \thirty\ , we get the value of $a$ as
$$ 1 - a = \pm { X \over K_0 }        \eqn\thirtyone  $$
where, using $K_0 = 2 N + 1 $ and $ \rho^2 = { {N(N^2 - 1)} \over {12} }$,
we get
$$ X = ( 1 + { {24} \over {N - 1} } G )^{ 1/2 } .  \eqn\thirtytwo $$
The equation \eighteen\ for $b_p , p = 1,2, \ldots, N-1 $
which parametrise $\vec \beta$ as
in \fifteen\ , remain the same and the solution is given by
$$ b_p = { K_0 \over 2 } + { {X l_p} \over {p (p + 1)} } \eqn\thirtythree $$
with $l_p $ defined as in \twenty\ . This gives the complete solution of
the spectrum of $W -$gravity satisfying the condition \thirty\ .
Now we further impose the requirement \twentynine\ where
$\epsilon_m^{\pm}$ are positive or negative integers only. Using \fifteen\
the equation \twentynine\ becomes
$$
K_+ \epsilon_m^+ + K_- \epsilon_m^- = m \sum_{p=m}^{N-1} b_p
\eqn\thirtyfour
$$
Note that the left hand side of \thirtyfour\ is an integer since all the
quantities appearing there are integers. The right hand side can be
written, taking the $ + $ sign in \thirtyone\ , as
$$
m \sum_{p=m}^{N-1} b_p = { m \over 2 } (K_0 - X) (N + 1)
- { {m (m + 1)} \over 2 } K_0 + X \sum_{p=1}^m i_p  .  \eqn\thirtyfive
$$
Consider the right hand side of the above equation. The second term is
always an integer. For the remaining sum to be an integer for all
$i_p$'s and $m$'s, $X$ must necessarily be an integer. However,this is
also a sufficient condition since  when $X$ is odd $(K_0 - X)$ is even and
when $X$ is even it can be seen
from \thirtytwo\ that $N$ must be odd. Hence the right hand side of
\thirtyfive\ is always an integer. Thus imposing \twentynine\ is
equivalent to requiring $X$ to be an integer.

The case $X = 1$ corresponds to the
canonical $W_N -$gravity spectrum with zero
ghost number. For this case, equations \thirtyfour\ and \thirtyfive\ imply
a very interesting phenomenon. Noting that $ \sum_{p=1}^m i_p $ takes the
values $ { 1 \over 2 } m (m + 1) + p , p = 0,1,2, \ldots ,m(N - m) $ , the
solutions for $\epsilon_m^{\pm} $  can be written as
$$
\epsilon_m^+ = p \, , \,  \epsilon_m^- = m (N - m) - p \, , \,
p = 0,1,2, \ldots ,m(N - m) . \eqn\insertone
$$
In particular
$ \sum_{m=1}^{N-1} ( \epsilon_m^+ + \epsilon_m^- ) =  (2 \rho^2) $ ,
the $L_0 -$intercept of the physical operators. Hence it follows that all
the physical operators in $W_N -$gravity with zero ghost number can be
expressed as the non singular composites of the screening charges. See \DDR\
for more details where this fact was first discovered for $N = 3,4$ and
conjectured for all $N $ .

Now, from \thirtythree\ the dimension $\Delta_p$ of
$e^{i \beta_p \phi_p} $ can be computed. It is given by
$$
\Delta_p =  { 1 \over { (2 \rho_p)^2 2 N (N + 1) } } \;
     ( ({ { K_0 p (p + 1) } \over 2 })^2 - X^2 l_p^2 ) . \eqn\thirtysix
$$
In particular, the dimension $\Delta_1 $ of
$e^{i \beta_1 \phi_1} $, where $\phi_1 $ is the ``liouville mode" , is given
by

$$
\Delta_1 \equiv 1 - h_1 (N) = 1 - { {X^2 l_1^2 - 1} \over {4 N ( N + 1)} } .
\eqn\thirtyseven
$$
As before, the range of $l_1 $ is $ 1 \leq \vert l_1 \vert \leq N-1 $ .
Comparing $h_1 (N)$
above with $h_{p,q} $ of $(N,N+1)$ minimal models we see that
the special operators of $W -$gravity constructed above can be interpreted
as the liouville dressings of the $(N,N+1)$ minimal model operators,
from both inside and outside the minimal table.
The fact that the new operators represent also the minimal model operators
from outside the minimal table is a welcome feature in the light of the
discovery \D\K\ that such minimal model operators fail to decouple
when coupled to 2d gravity.

Since the expression for the BRST charge Q$_B$ for $W_N -$gravity,
$N > 3$, is not known we are not able to construct
the full physical operators that include
the ghost sector with non trivial ghost number. However it is quite
likely that the special operators, with their $W-$gravity sector
constructed as above, exist and are part of the full physical spectrum
as we have seen explicitly in the case of $W_3 -$gravity,
indicating a deep relation between $W -$gravity theories and
minimal models coupled to 2d gravity.

We conclude by noting that there does not yet exist
a complete classification of $W -$gravity states
that include those with non trivial ghost numbers, analogous to that of
\LZ\ . Though this is a very interesting problem in its own
right, its solution may have to await a thorough understanding of
$W -$matter couplings to $W -$gravity. However, our explicit construction
of new states in section 3 appears to indicate that the Ising model
coupled to 2d gravity possesses a higher symmetry - $W_3$ symmetry - that
can be seen by representing the Ising model in a special way by the fields
$(\phi_2 ; b_3, c_3) $ .
We also need to understand better
the meaning and the origin of the selection rule described in section 4.
These aspects may shed light on the special states of $W -$gravity
theories constructed in this paper and the intriguing connection
between $W -$gravity and minimal models coupled to 2d gravity.
This may also help us understand better $W -$gravity itself and the origin
of $W_k -$constraints in the matrix models.

\vskip 1.0cm

\centerline{ACKNOWLEDGEMENTS}

It is a pleasure to thank Sumit Das and Avinash Dhar for many helpful
discussions and a critical reading of the manuscript.
I would also like to thank Sunil Mukhi and Ashoke Sen for
explaining their work and the reference \LZ\ .

\refout

\end